\documentclass[12pt]{article}
\usepackage{a4wide,epsfig,psfrag,amsmath,amssymb,cite,scalefnt}

\newcommand{\gsim}{\;\rlap{\lower 3.5 pt \hbox{$\mathchar \sim$}} \raise 1pt
 \hbox {$>$}\;}
\newcommand{\lsim}{\;\rlap{\lower 3.5 pt \hbox{$\mathchar \sim$}} \raise 1pt
 \hbox {$<$}\;}

\newcommand{\lmMh}{L_{\mu H}}
\newcommand{\lMhMt}{L_{Ht}}

\newcommand{\ep}{\epsilon}

\begin{document}

\title{\vskip-3cm{\baselineskip14pt
    \begin{flushleft}
      \normalsize SFB/CPP-09-64\\
      \normalsize TTP09-23
  \end{flushleft}}
  \vskip1.5cm
  Virtual three-loop corrections to Higgs boson production in 
  gluon fusion for finite top quark mass
}

\author{\small 
  Alexey Pak, 
  Mikhail Rogal,
  Matthias Steinhauser
  \\
  {\small\it Institut f{\"u}r Theoretische Teilchenphysik,
    Universit{\"a}t Karlsruhe (TH)}\\
  {\small\it Karlsruhe Institute of Technology (KIT)}\\
  {\small\it 76128 Karlsruhe, Germany}
}

\date{}

\maketitle

\thispagestyle{empty}

\begin{abstract}
In this letter we present the three-loop virtual corrections to the 
Higgs boson production in the gluon fusion channel where finite top
quark mass effects are taken into account. We perform an asymptotic
expansion and manage to evaluate five terms 
in the expansion parameter $M_H^2/M_t^2$. A good convergence is
observed almost until $M_H\approx 2M_t$.

\medskip


\end{abstract}

\thispagestyle{empty}


\newpage


\section{Introduction}

Among the main tasks of the CERN Large Hadron Collider (LHC) will be
the uncovering of the mechanism which provides particles with their
masses. A crucial role in this respect is assigned to the Higgs boson
whose discovery is awaited with great eagerness.

At LHC the Standard Model Higgs boson is mainly produced in the so-called
gluon fusion process where two gluons couple via a closed quark loop to the
Higgs boson. The leading order (LO) process for this channel has been evaluated 
in Refs.~\cite{Wilczek:1977zn,Ellis:1979jy,Georgi:1977gs,Rizzo:1979mf} 
and already almost 15 years ago also the 
next-to-leading order (NLO) corrections became 
available~\cite{Dawson:1990zj,Spira:1995rr}.
To this order the production cross section could be evaluated without any
assumption on the hierarchy between the mass of the quark in the loop,
the Higgs boson mass, and the partonic center-of-mass energy.

On the contrary, at next-to-next-to-leading order (NNLO) only quantum
corrections involving the top quark Yukawa coupling are available.
It has been performed
under the assumption that the top quark is much heavier than the
Higgs boson. In this limit it is suggestive to construct an effective
theory where the top quark is integrated out. The coefficient function
of the corresponding effective operator has been computed to NNLO in
Refs.~\cite{Chetyrkin:1997un,Steinhauser:2002rq} (see also
Ref.\cite{Kramer:1996iq}) and the 
production cross section has been evaluated in 
Refs.~\cite{Harlander:2000mg,Harlander:2002wh,Anastasiou:2002yz,Ravindran:2003um}.
Let us mention that recently the virtual contributions to the
NNNLO corrections have been completed~\cite{Chetyrkin:1997un,Baikov:2009bg}.

In this letter we provide the first results beyond the
effective-theory approach where three building blocks are required to NNLO:
virtual three-loop corrections to the $2\to1$ process, two-loop
corrections to the $2\to2$ process where next to the Higgs boson a
parton is radiated off, and one-loop corrections with radiation of two
additional partons.
We present results for the three-loop
virtual corrections to the process $gg\to H$
including finite top quark mass effects. Our results constitute a building
block for the NNLO corrections beyond the heavy top quark limit.

Let us mention that there are also results beyond the fixed-order
approximation. In particular, in Ref.~\cite{Catani:2003zt} large logarithms in
connection with soft gluon radiation have been resummed.
A step further has been taken in Ref.~\cite{Ahrens:2008nc}, where certain
$\pi^2$ terms have been resummed leading to a perturbative series which is
significantly better behaved as compared to the unresummed approach.
In Ref.~\cite{Marzani:2008az} the limit of high partonic center-of-mass
energies has been considered for the gluon-fusion process and an approximation
for the NNLO cross section has been derived which goes beyond the $M_t\to\infty$
result. 
Electroweak corrections have been considered in
Refs.~\cite{Actis:2008ug,Anastasiou:2008tj}. 
For recent numerical predictions of Higgs boson production in gluon fusion
both at the Tevatron and the LHC we refer to Ref.~\cite{deFlorian:2009hc}.

The remainder of the paper is organized as follows: in the next
section we briefly describe details of our calculation and the
Section~\ref{sec::results} contains our results and conclusions.


\section{Calculation}

The existing NNLO calculations to the Higgs boson production have been
performed in the framework of an effective theory where the top quark
has been integrated out. In this way effective
vertices\footnote{The effective coupling has even been computed to
  four- and five-loop order in 
  Refs.~\cite{Chetyrkin:1997un} and~\cite{Schroder:2005hy,Chetyrkin:2005ia},
  respectively.} 
are generated between the 
Higgs boson and two, three or four gluons. The number of
loops to be considered for the calculation of the 
cross section is effectively reduced by one, leading to virtual two-loop
$2\to1$, one-loop $2\to2$ and tree-level $2\to3$ corrections.

\begin{figure}[t]
  \centering
  \includegraphics[width=0.3\linewidth]{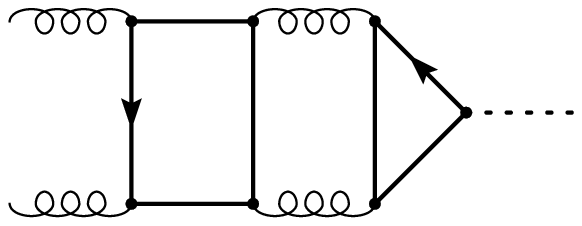}
  \put(-130,55){g}
  \put(-95,55){t}
  \put(-20,30){H}
  \hfill
  \includegraphics[width=0.3\linewidth]{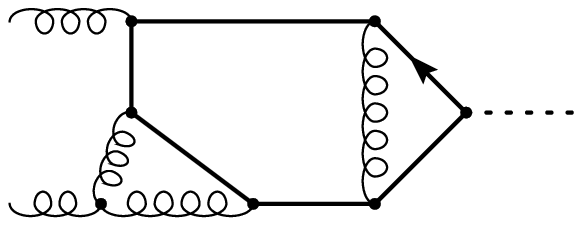}\hfill
  \includegraphics[width=0.3\linewidth]{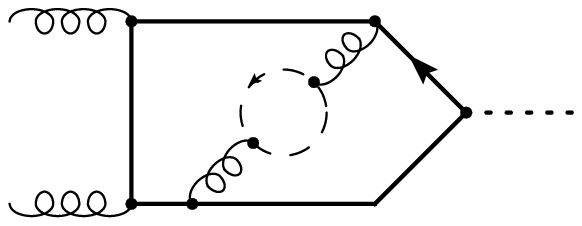}
  \put(-90,30){q}
  \caption[]{\label{fig::diag}Sample diagrams contributing to the NNLO virtual
    corrections to $gg\to h$.
    }
\end{figure}

In contrast to this approach, 
our starting point is the full QCD with six active flavours.
Some sample diagrams contributing to the virtual corrections 
are shown in Fig.~\ref{fig::diag}; altogether 657 three-loop diagrams have to be
considered which in the sum lead to the structure
\begin{eqnarray}
    X(\rho) (q_1\cdot q_2) g_{\mu\nu} 
    + Y(\rho) q_{1\nu} q_{2\mu} 
    + \ldots
  \,,
  \label{eq::Amunu}
\end{eqnarray}
where $\rho=M_H^2/M_t^2$ and 
$q_1$ and $q_2$ are the incoming momenta of the two gluons with
polarization vectors $\varepsilon^\mu(q_1)$ and $\varepsilon^\nu(q_2)$. In our
calculation we construct projectors on $X$ and $Y$ and check the condition
$X=-Y$ which follows from gauge invariance.
The ellipses in Eq.~(\ref{eq::Amunu}) represent further structures which do not
contribute to the physical cross section. They would receive contributions from
vertex corrections with external ghosts which we do not consider in this
paper.

The virtual contribution to the partonic 
cross section can be cast in the form
\begin{eqnarray}
  \hat{\sigma}_{ggh}^{\rm virt} &=& \hat{\sigma}_{\rm LO} \left(
    1 + \frac{\alpha_s}{\pi}~ \delta^{(1)}  
    + \left(\frac{\alpha_s}{\pi}\right)^2 \delta^{(2)} + \ldots
  \right)
  \,,
\end{eqnarray}
where the LO cross section is given by
\begin{eqnarray}
  \hat{\sigma}_{\rm LO} &=& \frac{G_F~\alpha_s^2}{288\sqrt{2}\pi} 
  \frac{f_0(\rho,\epsilon)}{(1-\epsilon)}~\delta(1-x) 
  \,,
\end{eqnarray}
with $x=M_H^2/\hat{s}$ and dimension of space $D = 4 - 2\ep$. 
$\sqrt{\hat{s}}$ is the partonic center-of-mass
energy and $f_0(\rho,\ep)$ reads
\begin{eqnarray}
  f_0(\rho,0) &=& \frac{36}{\rho^2}\left|1 + \left(1 -
      \frac{4}{\rho}\right)\arcsin^2
    \left(\frac{\sqrt{\rho}}{2}\right)\right|^2, 
  \qquad(\rho\le4)\,,
  \\ \nonumber 
  f_0(\rho,\epsilon) &=& \Bigg[1
    + \frac{7 + 7\ep}{60}\rho
    + \frac{1543 + 2486\ep + 943\ep^2}{100800}\rho^2
    + \frac{226 + 461\ep + 296\ep^2 + 61\ep^3}{100800}\rho^3 
    \nonumber \\
    &+& \frac{55354 + 130873\ep + 109848\ep^2 + 39533\ep^3 +
      5204\ep^4}{155232000}\rho^4 
    + \ldots
    \Bigg]
    \frac{\Gamma^2(1 + \ep)}{(M_t^2/\mu^2)^{2\epsilon}}
  \,,
\end{eqnarray}
where an expansion through $\mathcal{O}(\rho^4)$ and $\mathcal{O}(\ep^4)$ has been performed 
in the second line.
In Section~\ref{sec::results} we will present results for 
$\delta^{(1)}$ and $\delta^{(2)}$.

\begin{figure}[t]
  \centering
  \includegraphics[width=0.2\linewidth]{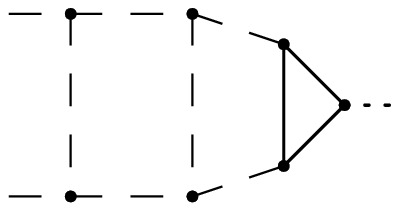}
  \put(-15,5){(a)}
  \hfill
  \includegraphics[width=0.2\linewidth]{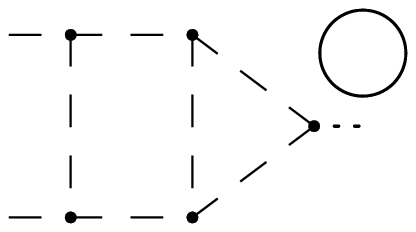}
  \put(-15,5){(b)}
  \hfill
  \includegraphics[width=0.2\linewidth]{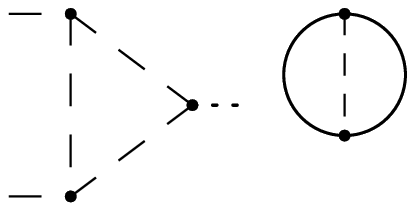}
  \put(-40,5){(c)}
  \hfill
  \includegraphics[width=0.2\linewidth]{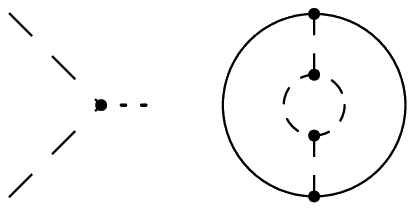}
  \put(-60,5){(d)}
  \caption[]{\label{fig::reg}Factorized regions appearing from the asymptotic 
    expansion of the double-scale integral (a). Solid lines carry the 
    mass $M_t$, dotted lines mass $M_H$, and the dashed lines are massless.
    Cases (b),(c), and (d) correspond to one, two, and three loop momenta at 
    the scale $M_t$, respectively, with the remaining loops at the scale $M_H$.}
\end{figure}

Note that the quantities $\delta^{(i)}$ only depend on $\rho$, since 
the Feynman diagrams have to be evaluated for on-shell external partons. 
It is thus tempting to evaluate them in the limit 
$M_H\ll 2M_t$ which is expected to show good convergence 
properties even up to $M_H\approx 2M_t$~\cite{Harlander:2003xy}. 
In Ref.~\cite{Schreck:2007um} the NNLO corrections to the decay of a Higgs boson
into gluons have been considered. The optical theorem in
combination with the asymptotic expansion was used in order to evaluate three expansion
terms in $M_H^2/M_t^2$ where rapid convergence has been observed for 
$M_H\approx M_t$.
The asymptotic expansion~\cite{Smirnov:2002pj} in the limit $M_H\ll 2M_t$
leads to one-, two- and three-loop vacuum integrals 
where the scale is given by the top-quark mass and to one- and two-loop 
vertex diagrams with massless internal lines and external momentum at the scale $M_H$.
In Fig.~\ref{fig::reg} we exemplify the asymptotic expansion in diagrammatic form 
for a typical contribution.

For our calculation we have used two independent set-ups. In the first one
all Feynman diagrams are generated with {\tt
  QGRAF}~\cite{Nogueira:1991ex}. The various
diagram topologies are identified and transformed to {\tt
  FORM}~\cite{Vermaseren:2000nd} notation with the help of {\tt q2e} and {\tt
  exp}~\cite{Harlander:1997zb,Seidensticker:1999bb}.  The program {\tt
  exp} is also used in order to apply the asymptotic expansion (see,
e.g., Ref.~\cite{Smirnov:2002pj}) in the various mass hierarchies. The
actual evaluation of the integrals is performed with the packages
{\tt MATAD}~\cite{Steinhauser:2000ry}, which is used for the vacuum
integrals, and {\tt FIRE}~\cite{Smirnov:2008iw}, employed to reduce
the massless three-point functions to master
integrals. The latter can, e.g., be found in Ref.~\cite{Gehrmann:2005pd}.

The second set-up also relies on {\tt QGRAF}
for the generation of the Feynman diagrams.
Afterwards the asymptotic expansion is done with a Perl program, and,
two- and three-loop integrals are reduced by an independent implementation
of the Laporta algorithm~\cite{Laporta:1996mq,Laporta:2001dd}.

We have performed the evaluation of the vertex corrections up to order
$\rho^2$ for general QCD gauge parameter and checked that it drops
out in the sum of all bare three-loop diagrams which serves as a welcome 
check of our calculation.

In the sum of all three-loop diagrams we observe poles up to order
$1/\epsilon^4$ which is due to a mixture of ultra-violet and infra-red
singularities. 
As usual, the ultra-violet poles are treated via renormalization. In our case
we have to renormalize the gluon wave function, top quark mass
and strong coupling constant to two-loop order.
The remaining infra-red poles are only cancelled after
including the corrections from the real radiation and mass factorization.

In the following section we present our results expressed in terms of
$\alpha_s^{(5)}$, 
the strong coupling constant in the $\overline{\rm MS}$ scheme 
defined in five-flavour QCD, and the on-shell top quark mass. 
Since our two-loop result contains poles up to ${\cal O}(1/\epsilon^2)$,
the one-loop top quark mass counterterm is needed up to 
${\cal O}(\epsilon^2)$ which can be found in Ref.\cite{Marquard:2007uj}.
The two-loop counterterm for $\alpha_s$ can be found, e.g., in
Ref.~\cite{Chetyrkin:2004mf} and the one for the gluon wave function
in Ref.~\cite{Chetyrkin:1997un}.

The transition from $\alpha_s^{(6)}$ to $\alpha_s^{(5)}$ is performed with the
help of the formulae derived in Ref.~\cite{Chetyrkin:1997un}.
Since there are poles in the NLO expression, higher
order terms in $\epsilon$ are necessary for the decoupling relation.
The explicit result can be found in Eq.~(12) of Ref.~\cite{Grozin:2007fh}.


\section{\label{sec::results}Results}

At the three-loop order we were able to evaluate the first five terms in the
expansion around $\rho = 0$. The NLO and NNLO corrections to the partonic cross
section are given by (adopting common $\overline{\rm MS}$ conventions)
\begin{eqnarray}
  \delta^{(1)} &=& -\frac{3}{\epsilon^2}
  + \frac{1}{\epsilon}\left(-\frac{23}{6} - 3 \lmMh\right)
  + \frac{11}{2} + \frac{21}{2}\zeta(2) - \frac{3}{2}\lmMh^2
  + \frac{34}{135}\rho 
  + \frac{3553}{113400}\rho^2 
  \nonumber\\&&
  + \frac{917641}{190512000}\rho^3 
  + \frac{208588843}{251475840000}\rho^4 
  + {\cal O}(\rho^5)
  \,,
\end{eqnarray}
\begin{eqnarray}
  \delta^{(2)} &=& \sum_{i\ge0} \delta^{(2)}_i \rho^i
  \,,
\end{eqnarray}
with
\begin{eqnarray}
  \delta^{(2)}_0 &=&
  \frac{9}{2\epsilon^4}
  + \frac{1}{\epsilon^3}\left(\frac{253}{16} + 9\lmMh\right)
  + \frac{1}{\epsilon^2}\left(-\frac{69}{8} - \frac{243}{8}\zeta(2)
    + \frac{115}{8}\lmMh + 9\lmMh^2 
  \right)
  \nonumber\\&&
  + \frac{1}{\epsilon}\left[
    -\frac{236}{9} 
    - \frac{621}{16}\zeta(2) - \frac{159}{8}\zeta(3)
    - \frac{33}{2}\lMhMt 
    + \frac{23}{4}\lmMh^2 + 6\lmMh^3 
  \right.\nonumber\\&&\left.\mbox{}
    + \lmMh\left(-\frac{943}{24} - \frac{243}{4}\zeta(2)\right) 
  \right]
  - \frac{125}{216} 
  + \frac{1547}{16}\zeta(2)
  + \frac{1161}{8}\zeta(4)
  - \frac{381}{8}\zeta(3)
  \nonumber\\&&
  - \frac{163}{8}\lMhMt
  - \frac{33}{4}\lMhMt^2
  +  3\lmMh^4 
  + \frac{23}{24}\lmMh^3
  + \lmMh^2\left(-\frac{943}{24} - \frac{243}{4}\zeta(2)\right)
  \nonumber\\&&
  + \lmMh\left(-\frac{1151}{72} 
    - \frac{69}{4}\zeta(2) 
    - \frac{159}{4}\zeta(3) 
    - 33\lMhMt 
  \right)
  \,,
\end{eqnarray}
\begin{eqnarray}
  \delta^{(2)}_1 &=&
    - \frac{34}{45\epsilon^2} 
    + \frac{1}{\epsilon}\left(-\frac{124997}{32400} 
      - \frac{34}{45}\lMhMt 
      - \frac{68}{45}\lmMh\right)
    - \frac{464570749}{10368000}
    - \frac{457253}{129600}\lMhMt
  \nonumber\\&&\mbox{}
    + \frac{211}{90}\zeta(2)
    + \frac{7}{45}\zeta(2)\ln2 
    + \frac{1909181}{55296}\zeta(3)
    - \frac{17}{45}\lMhMt^2
    - \frac{68}{45}\lmMh^2
  \nonumber\\&&\mbox{}
  + \left(-\frac{101537}{16200} - \frac{68}{45}\lMhMt\right)\lmMh 
  \,,
\end{eqnarray}
\begin{eqnarray}
  \delta^{(2)}_2 &=&
  - \frac{ 3553}{37800\epsilon^2} 
  - \frac{1}{\epsilon}\left(\frac{19652233}{38102400} 
    + \frac{3553 }{37800}\lMhMt 
    + \frac{3553 }{18900}\lmMh\right)
  -\frac{39974688999319}{4096770048000} 
  \nonumber\\&&\mbox{}
  +    \frac{887 }{3024}\zeta(2) 
  + \frac{857 }{37800} \zeta(2) \ln2 
  + \frac{267179777 }{35389440}\zeta(3)
  - \frac{24507239 }{50803200} \lMhMt
  - \frac{3553}{75600}  \lMhMt^2
  \nonumber\\&&\mbox{}
  - \frac{3553 }{18900}\lmMh^2 
  -  \lmMh\left(\frac{3244007}{3810240} 
    + \frac{3553 }{18900}\lMhMt \right) 
  \,,
\end{eqnarray}
\begin{eqnarray}
  \delta^{(2)}_3 &=&
 - \frac{917641}{63504000\epsilon^2} 
-\frac{1}{\epsilon}\left(\frac{ 13727463943}{160030080000 }
+ \frac{917641 }{ 63504000 }\lMhMt 
+ \frac{917641}{31752000  } \lmMh \right)
\nonumber\\&&\mbox{}
-\frac{12054084964483296871}{275302947225600000 } 
+ \frac{287809 }{6350400 } \zeta(2)
+ \frac{17881 }{4536000 }  \zeta(2)\ln2
  \nonumber\\&&\mbox{}
+ \frac{5756378217151 }{158544691200} \zeta(3)
 - \frac{5713528199 }{71124480000}\lMhMt
 - \frac{917641 }{127008000 } \lMhMt^2
- \frac{917641  }{31752000 } \lmMh^2
  \nonumber\\&&\mbox{}
-\lmMh\left( \frac{359730029 }{2500470000 } 
+\frac{ 917641}{31752000 }  \lMhMt \right)
    \,,
\end{eqnarray}
\begin{eqnarray}
  \delta^{(2)}_4 &=&
-\frac{ 208588843}{83825280000  \epsilon^2 }
- \frac{1}{\epsilon}\left(\frac{36471674738759}{2323636761600000 } 
+\frac{ 208588843 }{83825280000 } \lMhMt
  \right.\nonumber\\&&\left.\mbox{}
+\frac{ 208588843 }{41912640000 }\lmMh \right)
-\frac{749381165366796410981587}{21985693365436416000000 } 
 + \frac{65703703}{8382528000 }  \zeta(2)
  \nonumber\\&&\mbox{}
+ \frac{ 31270501 }{41912640000 }  \zeta(2) \ln2
+ \frac{89834770435139 }{3170893824000} \zeta(3)
 - \frac{45644737075181}{3098182348800000} \lMhMt
  \nonumber\\&&\mbox{}
- \frac{208588843 }{167650560000 } \lMhMt^2
- \frac{208588843 }{41912640000 } \lmMh^2
- \lmMh \left(
\frac{ 1933157007779 }{72613648800000 } 
  \right.\nonumber\\&&\left.\mbox{}
+ \frac{208588843 }{41912640000 } \lMhMt \right)
  \,,
  \label{eq::delta12}
\end{eqnarray}
where $\lmMh=\ln(\mu^2/M_H^2)$, $\lMhMt=\ln(M_H^2/M_t^2)$ and 
$\zeta(n)$ is the Riemann's zeta function. Furthermore, $SU(3)$ colour factors
have been applied and the number of massless quark flavours is set
to $n_l=5$. The analytic expression
for the generic values of $N_c$ and $n_l$ can be found in~\cite{ttpdata}.

\begin{figure}[t]
  \centering
  \begin{tabular}{cc}
    \includegraphics[width=.45\linewidth]{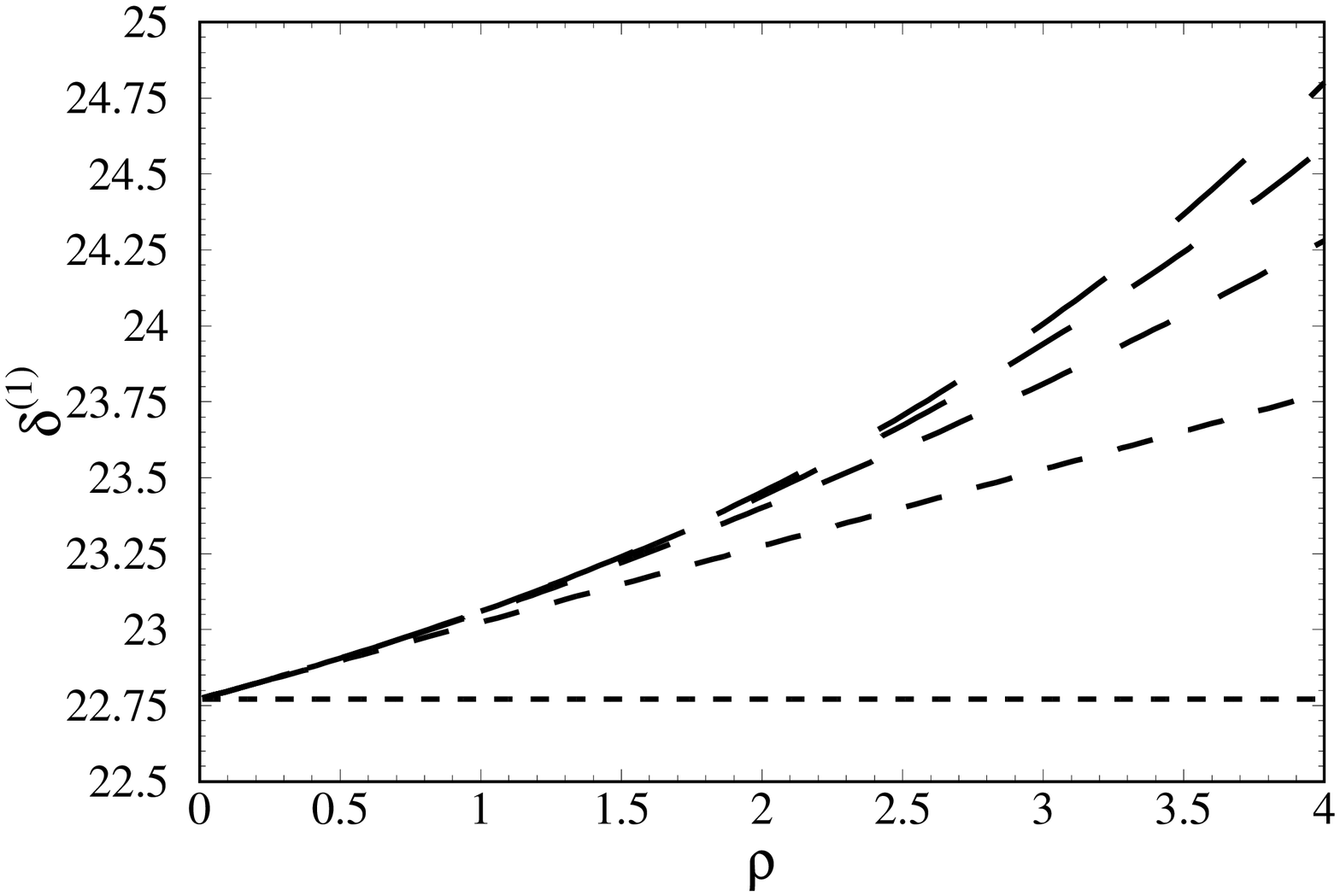}
    &
    \includegraphics[width=.45\linewidth]{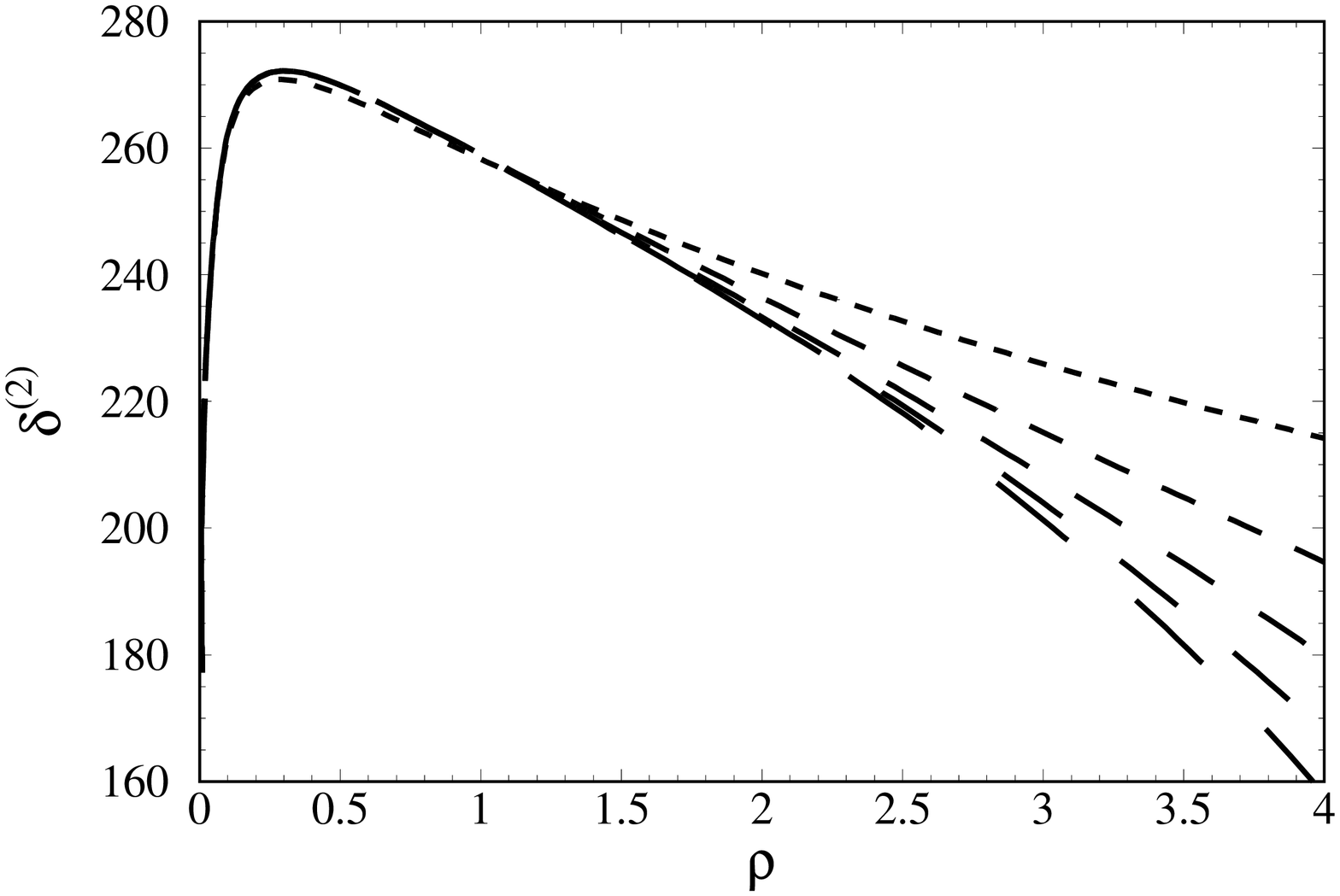}
  \end{tabular}
  \caption[]{\label{fig::del12}Finite part of $\delta^{(1)}$ (left) and $\delta^{(2)}$
    (right) as a function of $\rho$. The longer-dashed lines include
    successively higher orders in $\rho$.}
\end{figure}

We have checked that the expansion of $\delta^{(1)}$ to order $\rho$
agrees with~\cite{Dawson:1993qf}. At the NNLO the leading term in the
inverse top quark mass expansion agrees with the results of
Refs.~\cite{Chetyrkin:1997un,Harlander:2000mg}.

Although the final result is divergent, it is instructive to 
look at the finite parts of $\delta^{(1)}$ and $\delta^{(2)}$.
In Fig.~\ref{fig::del12} the corresponding two- and three-loop expressions are
shown for $\mu=M_H$ in the range between $\rho=0$ and $\rho=4$
corresponding to $M_H=2M_t$. 
The longer-dashed lines include successively higher orders in $\rho$
up to order $\rho^4$. 
One observes good convergence up to $\rho\approx 3$ which corresponds to
$M_H\approx 1.7 M_t$.

To conclude, we have presented the virtual corrections to the partonic cross
section $gg\to H$ including finite top quark mass effects. Our 
calculation confirms the results obtained 
in the framework of the effective theory and provides four
more expansion terms in $M_H^2/M_t^2$.
We observe a rapid convergence almost up to $M_H\approx 2 M_t$.
The results presented in this letter constitute a building block for the NNLO
corrections to the Higgs boson production in the gluon fusion channel
beyond the heavy top quark limit.

{\it
When this paper was in preparation, we had a chance to learn 
about the parallel publication~\cite{Harlander:2009} and establish the full
agreement of the results.
}



\vspace*{2em}
{\large\bf Acknowledgements}

We thank Johann K\"uhn for useful discussions and Robert Harlander and Kemal
Ozeren for providing us their results prior to publication.
This work was supported by the DFG through the SFB/TR~9 ``Computational
Particle Physics'' and by the BMBF through Grant No. 05H09VKE. M.R. was
supported by the Helmholtz Alliance ``Physics at the Terascale''. 







\end{document}